\documentclass{emulateapj}       
       
\shorttitle{Prompt emission in decaying magnetic field} \shortauthors{Pe'er \& Zhang}       
  
\newcommand{\cm}{\rm{\, cm}}      
\newcommand{\dL}{\rm{\, d_{L,28.5}^{-2}}}

\newcommand{\eV}{\rm{\, eV }}  
\newcommand{\keV}{\rm{\, keV }}       
\newcommand{\MeV}{\rm{\, MeV }}

\newcommand{\flu}{\rm{\, erg \, cm^{-2} \, s^{-1}}}       
       
\newcommand{\beq}{\begin{equation}}       
\newcommand{\eeq}{\end{equation}}       
\newcommand{\ba}{\begin{array}}       
\newcommand{\ea}{\end{array}}       
\renewcommand{\L}{L_{52}}    
    
\newcommand{\Gi}{\Gamma_{2}}  
\newcommand{\gm}{\gamma_{\min}}  
\newcommand{\gM}{\gamma_{\max}}    

\newcommand{\drb}{\Delta {r'}_{B,7}}

\newcommand{\ri}{r_{13}}
\newcommand{\ee}{\epsilon_{e,-0.5}}      
\newcommand{\eB}{\epsilon_{B,-0.5}}      
      
\newcommand{\epl}{\epsilon_{pl,0}}      
\def \etal{{\it et al.~}}      
      
\begin{document}       
    
\title{Synchrotron emission in small scale magnetic field as possible
  explanation for prompt emission spectra of Gamma-Ray bursts}
\author{Asaf Pe'er\altaffilmark{1} and Bing Zhang\altaffilmark{2}}
\altaffiltext{1}{Astronomical institute ``Anton Pannekoek'', Kruislaan
  403, 1098SJ Amsterdam, the Netherlands; apeer@science.uva.nl}
\altaffiltext{2}{Department of physics, University of Nevada, Las
  Vegas, NV 89154, USA}
 
\begin{abstract}
  Synchrotron emission is believed to be a major radiation mechanism
  during gamma-ray bursts' (GRBs) prompt emission phase.  A
  significant drawback of this assumption is that the theoretical
  predicted spectrum, calculated within the framework of the
  ``internal shocks'' scenario using the standard assumption that the
  magnetic field maintains a steady value throughout the shocked
  region, leads to a slope $F_\nu \propto \nu^{-1/2}$ below $ 100
  \keV$, which is in contradiction to the much harder spectra
  observed.  This is due to the electron cooling time being much
  shorter than the dynamical time.  In order to overcome this problem,
  we propose here that the magnetic field created by the internal
  shocks decays on a length scale much shorter than the comoving width
  of the plasma.  We show that under this assumption synchrotron
  radiation can reproduce the observed prompt emission spectra of the
  majority of the bursts.  We calculate the required decay length of
  the magnetic field, and find it to be $\sim 10^{4} - 10^{5}$~cm
  (equivalent to $10^5 - 10^6$ skin depths), much shorter than the
  characteristic comoving width of the plasma, $\sim 3 \times
  10^{9}$~cm.  We implement our model to the case of GRB050820A, where
  a break at $\lesssim 4 \keV$ was observed, and show that this break
  can be explained by synchrotron self absorption.  We discuss the
  consequences of the small scale magnetic field scenario on current
  models of magnetic field generation in shock waves.
\end{abstract}    

\section{Introduction}
\label{sec:into}

A widely accepted interpretation of the non-thermal radiation observed
during the prompt emission phase of gamma-ray bursts (GRBs) is that
synchrotron emission is a leading radiation mechanism during this
phase \citep{MLR93, MR93a, MR93b, MRP94, Katz94, RM94, Tavani96a}.
Indeed, early works found that the majority of bursts show spectral
slopes in the $\sim 1 - 200 \keV$ range of $\nu F_\nu \propto
\nu^{\alpha}$, with $\alpha \simeq 4/3$ \citep{Tavani96a, Tavani96b,
Cohen97, Schaefer98, Frontera00}, which is in accordance with the
predictions of the optically thin synchrotron emission model, provided
that the synchrotron cooling time of the radiating electrons is longer
than the emission time. In addition, recent comprehensive analysis of
the brightest BATSE bursts \citep{Preece00, Kaneko06} found that the
distribution of the low energy spectral slope peaks at $\alpha \simeq
1$, and a significant fraction of the bursts show spectral slope
consistent with $\alpha \simeq 4/3$.

The idea that synchrotron emission is the leading radiation mechanism
had gained further support by modeling the more detailed observations
of the afterglow phase in GRBs, which are found to be in good
agreement with this model prediction \citep{SNP96, MR97, W97a, W97b,
SPN98, PM98, WG99}.

In the standard internal/external shock scenario of GRBs \citep[the
``fireball'' model scenario;][]{RM92, RM94,SP97}, magnetic fields are
generated by shock waves. Electrons are accelerated to high energies
by the same shock waves, which thus provide the necessary conditions
for synchrotron radiation. The mechanisms of energy transfer to the
magnetic field and to accelerated electrons are not fully understood.
It is therefore common to parametrize the energy densities in the
magnetic field and in the energetic electrons as fractions
$\epsilon_B$ and $\epsilon_e$ of the post shock thermal energy, where
the values of $\epsilon_e$ and $\epsilon_B$ are inferred from
observations. By modeling GRB afterglow emission data \citep{WG99,
FW01, PK01, PK02}, the parameters values are found to be at least few
percents in most of the cases, and in some cases close to
equipartition \citep{WG99, Frail00}.

The fact that a significant fraction of the bursts show spectra that
are too hard to account for in the optically thin synchrotron model
\citep{Crider97, Preece98, Preece02, Ghirlanda03} had motivated works
on alternative emission models. These include Synchrotron self Compton
(SSC) scattering, first suggested by \citet{Liang97, LKSC97}, Compton
drag \citep{LGCR00}, upscattering of synchrotron self absorbed photons
\citep{GC99, PM00, Kumar06}, and Compton scattering of photospheric
photons \citep[][and references therein]{MR00, MRRZ02, PMR05, PMR06}.
While in principle these models can reproduce a hard spectral slope,
common requirements to all models involved inverse Compton scattering
as a leading radiation mechanism is that at the emission radius the
optical depth to scattering is high, and that $\epsilon_B/\epsilon_e
\ll 1$. This last requirement, in turn, can lead to extensive
radiation at very high ($\gg \MeV$) energies, thus to low radiative
efficiency at the sub-MeV energy range \citep{DKK01}.  Additional
drawback of SSC models is the wider spread in the peak energy
distribution (compared to synchrotron model results for similar range
of parameter dispersion), which might not be consistent with the data
\citep[e.g.,][]{ZM02}.  Therefore, these models put various
constraints on the allowed parameter space region during the emission
phase \citep[e.g.,][]{ZM04, PW04}.

An argument raised against the synchrotron mechanism [\citet{GCL00};
 see also discussion in \citet{ZM04}] is that the inferred values of
 the free model parameters, in particular the strength of the comoving
 magnetic field during the prompt emission phase, $B' \sim 10^5 -
 10^6$~G, implies that the radiating electrons are synchrotron cooled
 much faster than the dynamical time.  This, in turn, leads to a
 spectrum with slope $\nu F_\nu \propto \nu^{1/2}$ below $\sim 100
 \keV$, which is in conflict to the much harder spectra observed in
 this energy range.  In order to overcome this problem, it was
 suggested that the energy distribution of radiating electrons has a
 smooth cutoff, and that the pitch angles of these electrons are
 anisotropically distributed \citep{LP00, LP02}.

A crucial underlying assumption in this analysis, is that electrons
radiate on a length scale comparable to the entire comoving width of
the shocked plasma. For plausible assumptions about the number
density and characteristic Lorentz factors in GRBs, this assumption
can only hold if the magnetic field maintains approximately constant
value on a scale of $\approx 10^9$ skin depth \citep{Piran05}.

Generation of magnetic field in strong, relativistic shock waves is
still poorly understood. Two stream instability of flow past shock
waves can, in principle generate strong magnetic field
\citep{ML99}. However, state of the art numerical models
\citep{Silva03, Fred04, Nishikawa05} can only trace the evolution of
this field on a characteristic scale of few tens of skin depths at
most, due to the huge numerical effort involved. The evolution of the
magnetic field on larger scale therefore still remains an open
question.  While some models predict that the magnetic field saturates
at a value close to equipartition \citep[e.g.,][]{JLT04}, several
authors find much weaker magnetic field \citep{WA04}, or argued that
the created magnetic field quickly decays by phase-space mixing
\citep{Gruz01}.

Motivated by these uncertainties on the length scale of the magnetic
field, \citet{RR03} suggested a model for GRB afterglow emission, in
which the magnetic field decays on a length scale shorter than the
shocked region scale. In that work, however, the decay length of the
magnetic field was not specified. 

In this paper we show that by assuming that the magnetic field decays
on a length scale shorter than the comoving scale, the observed prompt
emission spectra of many GRBs can be reproduced.  Thereby, this
assumption allows to overcome the ``fast cooling time'' problem
inferred by \citet{GCL00}.  We calculate in section
\S\ref{sec:calculations} the values of the free model parameters that
can account for the GRBs prompt emission spectra. We show that the
decay length of the magnetic field that is consistent with the
observed spectra is $\sim 10^4 - 10^5$~cm, which is $\approx 10^{5.5}$
skin depths.  We then implement our model in section
\S\ref{sec:GRB050820A} to the specific case of GRB050820A, where a low
energy break at $\sim 4 \keV$ was observed. We summarize our results
and discuss the implications of our model in view of current models of
magnetic field generation in relativistic shock waves in section
\S\ref{sec:summary}.

\section{Theory of small magnetic field length scale: constraints on
  model parameters set by observations}
\label{sec:calculations}

We adopt the framework of the internal shock scenario and assume that
variability in the Lorentz factor $\Gamma$ of the relativistic wind
emitted by the GRB progenitor leads to the formation of shock waves
within the expanding wind at radii much larger than the underlying
source size \citep[see, e.g.,][]{ZM02}. We assume that these shock
waves, produced at characteristic radius $r$ from the progenitor, are
the source of the magnetic field.  We introduce a new length scale
$\Delta r'_B$, which is the comoving length scale characterizing the
decay of the magnetic field. This decay length is much shorter than
the comoving width of the plasma $\Delta r' \simeq r/\Gamma$.  We
derive in this section the constraints on the model parameters as
inferred from observations.

The radiating electrons are accelerated by the shock waves to a power
law distribution with power law index $p$ above some characteristic
energy $\gamma_{\min} m_e c^2$. Synchrotron radiation by these
electrons is the main emission mechanism, therefore the break energy
observed in many bursts at $\varepsilon_m^{ob.} \gtrsim 100\keV$ is
attributed to synchrotron radiation from electrons at $\gamma_{\min}$.
Denoting by $\gamma_c$ the Lorentz factor of electrons that cool on a
time scale equal to the dynamical timescale, and by
$\varepsilon_c^{ob.}$ the characteristic observed energy of photons
emitted by synchrotron radiation from these electrons, the requirement
that the spectral slope $\nu F_\nu \propto \nu^{\alpha}$ has a
characteristic spectral index $\alpha \simeq 4/3$ below $\sim 100
\keV$ leads to $\varepsilon_c^{ob.}  \gtrsim 100\keV$. The value of
$\varepsilon_c^{ob.}$ can not be much greater than $100 \keV$, in
order to ensure high radiative efficiency (see discussion in
\S\ref{sec:summary} below).

The requirement that the spectral slope is not harder than $4/3$, as
is the case in a significant fraction of the bursts, implies that in
these bursts inverse Compton scattering and thermal emission component
do not play a significant role in producing the spectra below $100
\keV$. These conditions can be translated to the demand that the
emission radius $r$ is larger than the photospheric radius, $r_{ph}$.
Additional two constraints are that the observed flux $\nu F_{\nu}$
and the synchrotron self absorption energy $\varepsilon_{ssa}^{ob.}$,
which produces a low energy break, are consistent with observations.

The observational constraints can therefore be written as a set of
equations in the form:
\beq
\ba{llll}
(a) & \varepsilon_m^{ob.} & \gtrsim  100 & \keV, \nonumber \\
(b) & \varepsilon_c^{ob.} & \gtrsim  100 & \keV, \nonumber \\
(c) & r & \gtrsim r_{ph} & , \nonumber \\
(d) & \nu F_{\nu}^{ob.} & \simeq  10^{-7}& {\rm \,
  erg\,s^{-1}\,cm^{-2}}, \nonumber \\
(e) & \varepsilon_{ssa}^{ob.}  & \lesssim 1 & \keV,
\label{eq:obs}
\ea
\eeq  

We now apply the set of equations (\ref{eq:obs}) describing the
constraints set by observations to constraints on the uncertain values
of the free model parameters.

Assuming variability in the Lorentz factor $\Delta \Gamma/ \Gamma \sim
1$ on timescale $\Delta t$ of the expanding relativistic wind, shocks
develop at radius $r \simeq \Gamma^2 c \Delta t$. Due to Lorentz
contraction, the comoving width of a plasma shell is $\Delta r' \simeq
r/\Gamma$.  We use the standard fireball model assumption, in which
the burst explosion energy is initially converted to
kinetic energy. For isotropically equivalent central engine luminosity
$L$ which is time independent over a period $\Delta t$, the
isotropically equivalent number of protons ejected from the progenitor
during this period is $N_p \approx L \Delta t / \Gamma m_p c^2 $.
Therefore, the comoving number density of protons in the shock heated
plasma is given by
\beq 
n_p'(r) \approx {\zeta L \over 4 \pi r^2 \Gamma^2 c m_p
c^2} = 1.8 \times 10^{13} \, \L \, \ri^{-2} \, \Gi^{-2} \, \zeta_0 \,
{\rm cm^{-3}},
\label{eq:n}
\eeq
where $\zeta$ is the
compression ratio ($\zeta \simeq 7$ for strong shocks) and the
convention $Q = 10^x Q_x$ is adopted in CGS units.  Assuming that the proton
internal energy (associated with the random motion) in the shocked
plasma is $\theta_p m_p c^2$, the comoving internal energy density is
$u' = n_p' \theta_p m_p c^2$. The value of $\theta_p$ is not expected to
be much larger than {\it a few} at most for mildly relativistic (in
the comoving frame) shock waves.  The magnetic field carries a
fraction $\epsilon_B$ of the internal energy density, thus the comoving
magnetic field strength is given by
\beq
\ba{lll}
B' & = & \sqrt{8 \pi \epsilon_B u'} \nonumber \\
& = & 4.6 \times 10^5 \, \L^{1/2} \,
\ri^{-1} \, \Gi^{-1} \, \eB^{1/2} \, \theta_{p,0}^{1/2} \,
\zeta_0^{1/2} {\rm \, G}. 
\ea
\label{eq:B}
\eeq 

We assume that a fraction $\epsilon_{pl} \leq 1$ of the electron
population is accelerated by the shock waves to a power law energy
distribution with power law index $p$ above $\gamma_{\min}$ (and below
$\gamma_{\max}$).  Assuming that a fraction $\epsilon_e$ of the
post-shock thermal energy is carried by these electrons, the minimum
Lorentz factor of the energetic electrons is given by
\beq
\ba{lll}
\gamma_{\min} & = & {\epsilon_e \theta_p \over \epsilon_{pl}} 
\left( {m_p \over m_e} 
\right) {1 - (\gamma_{\min}/\gamma_{\max}) \over 
  \log{\left({\gamma_{\max} \over \gamma_{\min}}\right)} } \times
\Psi(p) \nonumber \\ 
& = & 86 \, \ee \, \epl^{-1} \, \theta_{p,0}  \, \Psi(p),
\ea
\label{eq:gm}
\eeq 
where characteristic value $\log(\gamma_{\max}/\gamma_{\min})
\simeq 7$ was used. The function $\Psi(p)$ determines the dependence
of the value of $\gamma_{\min}$ on the power law index $p$ of the
accelerated electrons, and is normalized to $\Psi(p=2) = 1$. A full
calculation of this function for various values of the power law index
$p$ is given in appendix \ref{psi_calc}.  
Using equations \ref{eq:B} and \ref{eq:gm}, the break in the
spectrum from burst at redshift $z$ is observed at
\beq
\ba{lll}
\varepsilon_m^{ob.} & = & \left({1 \over 1 + z}\right){3 \over 2} \hbar
\Gamma { q B' \gamma_{\min}^2 \over  m_e c}  \nonumber \\ 
& = & 
 {5.9 \over 1 + z} \, \L^{1/2} \, \ri^{-1} \, \ee^2 \, \epl^{-2} \,
\eB^{1/2} \nonumber \\  
& & \times \theta_{p,0}^{5/2} \, \zeta_0^{1/2} \, \Psi^2(p) \keV .
\ea
\label{eq:em}
\eeq 

Electrons in the shocked region propagate at velocity close to the
speed of light. Therefore, electrons  cross 
the magnetized area in a comoving time $\approx \Delta r'_B/c$. Since
this is the 
available time for electrons to radiate, equating the synchrotron
cooling time and the crossing time of this area gives the cooling
break of the electrons energy distribution, which occurs at Lorentz
factor $\gamma_c = (9 m_e^3 c^6)/(4 q^4 B'^2 \Delta r'_B)$. 
Photons emitted by electrons at $\gamma_c$ are observed at energy
\beq
\varepsilon_c^{ob.} =  {97 \over 1 + z} \, \L^{-3/2} \, \ri^3 \, \Gi^4
\, \eB^{-3/2} \, \drb^{-2} \, \theta_{p,0}^{-3/2} \, \zeta_0^{-3/2} \eV. 
\label{eq:ec}
\eeq 

The number of radiating electrons is calculated by integrating the
number density of energetic electrons inside the emitting region,
$N_e(r) = 4 \pi \int_r^{r+\Delta r_B} r^2 n(r) dr \simeq (\zeta
\epsilon_{pl} L / \Gamma m_p c^3)(\Delta r'_B /\Gamma)$. Here, $\Delta
r_B = \Delta r'_B/\Gamma$ and $n(r) = \Gamma \epsilon_{pl} n'_p(r)$
are the (observer frame) width and number density of radiating
electrons inside this region
\footnote{Due to the requirement $r > r_{ph}$, a significant number of
pairs cannot be created.}.

By requirement, $\gamma_c \geq \gamma_{\min}$, therefore in
calculating the observed flux, one can approximate the photon energy
to be close to $\varepsilon_m^{ob.}$. The (frequency integrated) power
emitted by electrons with Lorentz factor $\gamma_{\min}$ is
$P(\gamma_{\min}) = (4 q^4 B'^2 \gamma_{\min}^2)/(9 m_e^2 c^3)$,
therefore the observed flux is
\beq
\ba{lcl}
\nu F_{\nu}^{ob.} & = & { P(\gamma_{\min}) N_e(r) \over 4 \pi d_L^2} \times
\Gamma^2 \nonumber \\
& = & 2.9 \times 10^{-7} \, \L^2 \, \ri^{-2} \, \ee^2 \, \epl^{-1} \,
\eB \nonumber \\ 
& & \times \drb \, \Gi^{-2} \dL \, \theta_{p,0}^{3} \, \zeta_0^{2} \,
\Psi^2(p) \,\flu,
\ea
\label{eq:F}
\eeq 
where $d_L=10^{28.5} d_{L,28.5} \cm$ is the luminosity distance, and a
factor $\Gamma^2$ is introduced to transform the result from the
comoving frame to the observer frame.

The optical depth is given by $\tau(r) = n'_p(r) \Delta r' \sigma_T$,
where the comoving width $\Delta r'$ and not the comoving radiating
width $\Delta r'_B$ appears in the equation since electrons scatter
photons outside the radiating region as well. The photospheric radius
is thus given by
\beq
\ba{lll}
r_{ph} = r(\tau(r)=1) & = & { \zeta L  \sigma_T \over 4 \pi \Gamma^3
  m_p c^3}  \nonumber \\
& = & 1.2 \times 10^{13} \, \L \, \Gi^{-3} \, \zeta_0 \, \cm.
\ea
\label{eq:r_ph}
\eeq

The observed synchrotron self absorption energy break is calculated
using standard formula \citep[e.g.,][]{Rybicki79}  
\beq
\ba{lll}
\varepsilon_{ssa}^{ob.} & = & {31 \over 1 + z} \, \L^{4/5} \, \ri^{-8/5}
\, \drb^{3/5} \, \Gi^{-3/5} \, \ee^{-1} \, \epl^{8/5} \, \eB^{1/5} 
\nonumber \\
& & \times \theta_{p,0}^{-4/5} \, \zeta_0^{4/5} \chi(p) \, \eV,
\ea
\label{eq:e_ssa}
\eeq
where $\chi(p)$ is a function of the power law index $p$ of the
accelerated electrons, which is normalized to $\chi(p=2)=1$.  We
present in appendix \ref{chi_calc} a full derivation of this function,
and show that its value strongly depends on the uncertain value of the
power law index $p$ of the accelerated electrons.
       
While the first four constraints in equation \ref{eq:obs} (a-d) are
common to the majority of bursts, observation of a low energy break,
which may be attributed to synchrotron self absorption frequency is 
controversial. We therefore treat the last constraint in equation
\ref{eq:obs} separately in section \S\ref{sec:GRB050820A}.

The constraints set by observations in equation \ref{eq:obs} (a-d) can
be written with the use of equations (\ref{eq:em})-(\ref{eq:r_ph}) in
the form
\beq
\ba{llcr}
(a) & \left({5.9 \over 1+z }\right) \,
\L^{1/2} \, \ri^{-1} \, \ee^2 \, \epl^{-2} \, \eB^{1/2} \,
\theta_{p,0}^{5/2} \, \zeta_0^{1/2} \, \Psi^2(p) \nonumber \\ 
 & \qquad \qquad \qquad  \; \, = 100 \, \alpha_1, \nonumber \\  
(b) & \left({0.097 \over 1 + z}\right) \,
\L^{-3/2} \, \ri^3 \, \drb^{-2} \, \Gi^4 \, \eB^{-3/2} \, 
\theta_{p,0}^{-3/2} \, \zeta_0^{-3/2} \nonumber \\ 
 & \qquad \qquad  \qquad  \; \, =  100 \, \alpha_2, \nonumber \\
(c) & \ri \qquad \qquad \; \; \; =  1.2 \, \L \, \Gi^{-3} \, \zeta_0 \, \alpha_3,
\nonumber \\ 
(d) & 2.9  \, \L^2 \, \ri^{-2} \, \drb \, \Gi^{-2} \, \ee^2 \,
\epl^{-1} \, \eB \, \dL  \nonumber \\ 
& \times \theta_{p,0}^{3} \, \zeta_0^{2} \,
\Psi^2(p) = 1 \, \alpha_4, \nonumber \\ 
\ea
\label{eq:const1}
\eeq
where the free parameters $\alpha_1 - \alpha_4$ are introduced
in order to replace the inequalities in equation \ref{eq:obs} by
equalities, thereby account for the variety of GRB data.

In order to derive constraints on the values of the free model
parameters from the set of equations \ref{eq:const1} (a-d), we note
that the parameters $\epsilon_e$ and $\epsilon_B$ are constrained
from above by a maximum allowed value of equipartition ($\ee , \eB
\leq 1$). The parameter $\epsilon_{pl}$ also has an upper limit,
$\epsilon_{pl} \leq 1$. Furthermore, the values of $\theta_p$, $\zeta$
and $\Psi(p)$ (for $p \geq2$) can only be larger or equal to unity. 
In contrast to these constraints, there are no further intrinsic
constraints on the values of the isotropic equivalent luminosity
$L$, the emission radius $r$, the fluid Lorentz factor $\Gamma$ and
the comoving decaying length of the magnetic field $\Delta r'_B$. 
We therefore solve the set of equations \ref{eq:const1}
(a-d) to find the values of $L$, $r$, $\Gamma$ and $\Delta r'_B$, and
obtain
\beq
\ba{lcll}
\L & = & 2.7  \; \; d_{L,28.5}^{2} \, \ee^{-1} \, \theta_{p,0}^{-1} \,  
\zeta_0^{-1} \, \Psi^{-1}(p) \nonumber \\ 
& & \times \alpha_1^{-1/2} \, \alpha_2^{1/2} \,
\alpha_4, \nonumber \\
\ri & = & 0.1 \, (1+z)^{-1} \; \;  d_{L,28.5} \, \ee^{3/2} \, \epl^{-2} \,
\eB^{1/2} \nonumber \\
& & \times \theta_{p,0}^{2} \,  \Psi^{3/2}(p) \, 
 \alpha_1^{-5/4} \, \alpha_2^{1/4} \, \alpha_4^{1/2}, \nonumber \\
\drb & = & 4.6 \times 10^{-3} \, (1+z)^{-4/3} \; \; d_{L,28.5}^{2/3} \,
\ee^{4/3} \, \epl^{-5/3} \nonumber \\
& & \times \eB^{-1/3} \, \theta_{p,0} \,
\Psi^{4/3}(p) \, \alpha_1^{-1} \, \alpha_2^{-1/3} \, \alpha_3^{2/3} \,
\alpha_4^{1/3}, \nonumber \\ 
\Gi & = & 3.2 \, (1+z)^{1/3} \; \; d_{L,28.5}^{1/3} \, \ee^{-5/6} \,
\epl^{2/3} \, \eB^{-1/6} \nonumber \\
& & \times \theta_{p,0}^{-1} \,  \Psi^{-5/6}(p) \, 
 \alpha_1^{1/4} \, \alpha_2^{1/12} \, \alpha_3^{1/3} \, \alpha_4^{1/6}.
\ea
\label{eq:results1}
\eeq

The values of the free model parameters derived in equation
\ref{eq:results1} indicate that the prompt emission spectra of the
majority of the bursts can be explained in the framework of the model
suggested here. For values of $\epsilon_e$ and $\epsilon_B$ not far
below equipartition and $\epsilon_{pl}$ close to unity, these results
imply that the emission radius 
should be $r\lesssim 10^{12}$~cm, and that the magnetic field decays
on a comoving scale $\Delta r'_B \sim 10^{4.5}$~cm. If only $\approx
10\%$ of the electrons are accelerated in the shock waves,
$\epsilon_{pl} = 0.1$, then the emission radius is significantly
higher, $r\simeq 10^{14}$~cm, and the magnetic field decays after
$\Delta r'_B \sim 10^{6.5}$~cm.  Interestingly, the derived values of
the isotropically equivalent luminosity and the characteristic fluid
Lorentz factor are not different than their derived values in the
standard internal shock scenario. We further discuss the implications
of these results in \S\ref{sec:summary}.

\section{Possibility of a low energy break: The case of GRB050820A}
\label{sec:GRB050820A}

The results obtained in the previous section in equation
\ref{eq:results1}, may be applicable to many GRBs that show spectral
slope $\nu F_\nu \propto \nu^{4/3}$ below $\sim 100 \keV$.  For the
majority of GRBs observations during the prompt emission phase are
available only above few $\keV$ (BATSE, Beppo-SAX or SWIFT BAT-XRT
energy range). In most cases, observations do not indicate an
additional low-energy break in the spectrum that might be attributed
to synchrotron self absorption.  On the contrary, in some cases
\citep[e.g., GRB060124;][]{Romano06} interpolation of data taken in
the UV band supports the lack of an additional spectral break above
$\sim 1 \eV$.

Even though uncommon to many GRBs, an additional, second low energy
break may have been observed in some bursts. In at least one case -
GRB050820A \citep{Page05}, there are indications for a low energy
break at $\lesssim 4 \keV$ observed during a gamma-ray/X-ray giant
flare that occurred 218 seconds after the burst trigger, and lasted 34
seconds \citep{Osborne06}. This low energy break may be attributed to
synchrotron self absorption.

In order to account for these results in the framework of the model
presented here, we insert the values of the four parameters $L$, $r$,
$\Delta r'_B$ and $\Gamma$ as inferred from the observational
constraints in equation \ref{eq:results1}, to the equation describing
the observed self absorption frequency (equation \ref{eq:e_ssa}). This
insertion results in
\beq
\ba{lll}
\varepsilon_{ssa}^{ob.} & = & 56  \, (1+z)^{-2/5} \, d_{L,28.5}^{1/5} \,
\ee^{-29/10} \, \epl^{17/5} \,
\eB^{-7/10} \nonumber \\
& & \times \theta_{p,0}^{-18/5} \,  \tilde{\chi}(p) \, 
 \alpha_1^{17/20} \, \alpha_2^{-1/4} \, \alpha_3^{1/5} \,
 \alpha_4^{1/10} \, \eV ,
\ea
\label{eq:e_ssa2}
\eeq
where $\tilde{\chi}(p)$ gives the dependence of
$\varepsilon_{ssa}^{ob.}$ on the electrons power law index $p$, and is
normalized to $\tilde{\chi}(p=2)=1$.  We present in appendix
\ref{chi_calc} a full derivation of this function, and show there that
for values of $p$ in the range $2 \leq p \leq 2.4$, this function
varies by a factor less than 4.  We can thus conclude that within the
framework of the model suggested here, the self absorption energy is
not very sensitive to the uncertain value of the power law index $p$
of the accelerated electrons.

While the the self absorption break calculated in equation
\ref{eq:e_ssa2} is clearly lower than the value of the break energy
observed in GRB050820A, this equation indicates a very strong
dependence of the break energy on the uncertain values of the
parameters $\epsilon_e$, $\epsilon_{pl}$ and $\theta_p$. The
equipartition value of $\epsilon_e$ used in equation \ref{eq:e_ssa2}
is an upper limit. If the value is $\epsilon_e \approx 0.1$, then
the self absorption break is observed at $\sim 2 \keV$. Similarly, for
$\epsilon_{pl} \simeq 0.3$ or $\theta_p \approx 3$ , the self
absorption break occurs at $\sim 1 \eV$. Using the results of equation
\ref{eq:results1} we find that the values of the four other parameters
$L$, $r$, $\Gamma$ and $\Delta r'_B$ are much less sensitive to the
uncertainties in $\epsilon_e$, $\epsilon_{pl}$ or $\theta_p$.

The parameters $\epsilon_e$, $\epsilon_{pl}$ and $\theta_p$
parametrize the post shock energy transfer to the electrons, the
fraction of the electrons population accelerated by the shock waves,
and the normalized mean random energy gained by proton population.
All these physical quantities depend on the microphysics of energy
transfer and particle acceleration in shock waves, both of which are
not fully understood. We cannot therefore, from a theoretical point of
view, rule out the possibility that the values of $\epsilon_e$,
$\epsilon_{pl}$ and $\theta_p$ are sensitive to the plasma conditions
at the shock forming region.  Equation \ref{eq:e_ssa2} combined with
measurement (or constraint) on the self absorption frequency, can be
used to constrain the uncertain values of these parameters.

\section{Summary and discussion}
\label{sec:summary}

In this work we have presented a model in which the magnetic field
produced by internal shock waves in GRBs decays on a short length
scale. Using this assumption, we showed that the prompt emission
spectra of the majority of GRBs can be explained as due to synchrotron
radiation from shock accelerated electrons. We found that the required
(comoving) decay length of the magnetic field is $\sim 10^{4.5}$~cm,
and that the radiation is produced at $\sim 10^{12}$~cm from the
progenitor (eq. \ref{eq:results1}).  These parameter values were found
to be relatively sensitive to the fraction of electrons population
accelerated by the shock waves $\epsilon_{pl}$, and can therefore be
higher.  We showed in \S\ref{sec:GRB050820A} that the observed
synchrotron self absorption energy is very sensitive to the uncertain
values of the post shock thermal energy fraction carried by the
electrons, to the mean proton energy $\theta_p$ and to the value of
$\epsilon_{pl}$, thereby argued that the energy of a low energy break
is expected to vary between different bursts.

A major result of this work is the characteristic decay length of the
magnetic field deduced from observations, $\sim 10^{4.5}$~cm.  This
value is significantly shorter than the standard assumption used so
far, that the magnetic field strength is approximately constant
throughout the comoving plasma width, $\approx 10^{10}-10^{11} $~cm.
Still, the electron crossing time of the magnetized region is long
enough to allow electrons acceleration to high energies. Equating the
electron acceleration time, $t_{acc} \simeq \gamma m_e c^2 /(cqB')$
and the electron crossing time, $\Delta r'_B / c$, gives an upper
limit on the electron Lorentz factor, $\gamma_{\max,1} = (\Delta r'_B
q B')/(m_e c^2) \simeq 2 \times 10^7 \, \Delta r'_{B,4.5} B'_6$, where
$B' = 10^6 B'_6$~G.  This value is larger than the maximum electron
Lorentz factor obtained by equating the acceleration time and the
synchrotron cooling time, $\gamma_{\max,2} = (3/2) (m_e c^2)/(q^3
B')^{1/2} \simeq 10^5 {B'_6}^{-1/2}$. We thus conclude that within the
magnetized region, for $\epsilon_B \gtrsim 10^{-3}$ an upper limit on
the accelerated electron energy is set by the synchrotron cooling
time, and not by the physical size of this region.

The parameters $\alpha_1 - \alpha_4$ introduced in equations
\ref{eq:const1}, \ref{eq:results1} account for the difference between
the variety of GRB data and the characteristic values considered in
the analytical analysis.  As a concrete example, cooling energy
$\epsilon_c^{ob.}$ larger than $100 \keV$ is accounted for by
considering $\alpha_2 > 1$, which implies through equation
\ref{eq:results1} that high isotropic equivalent luminosity is
required in order to account for an observed flux $\simeq 10^{-7} {\rm
\, erg\,s^{-1}\,cm^{-2}}$.  This result is understood as due to the
low radiative efficiency in the case $\epsilon_c^{ob.} \gg
\epsilon_m^{ob.}$.

The ratio found here between the decay length of the magnetic field
and the comoving shell thickness,
\beq
\ba{lll}
{\Delta {r'}_B \over (r/\Gamma)} & = & 1.5 \times 10^{-5} \, \ee^{-1} \,
\epl \, \eB^{-1} \, \theta_{p,0}^{-2} \,  \Psi^{-1}(p) \nonumber \\
& & \times
\alpha_1^{1/2} \, \alpha_2^{-1/2} \, \alpha_3, 
\ea
\eeq
is based on fitting the GRB prompt emission spectra.  In earlier work,
based on modeling afterglow emission \citep{RR03}, this value was
thought to be too low, due to the high ambient medium density it
implies during the afterglow emission phase.  However, in the work by
\citet{RR03} detailed modeling of afterglow data was not performed,
due to lack of available data. Moreover, the well established
connection between long GRBs and core collapse of massive stars
\citep[e.g.,][and references therein]{PW06} indicates that indeed the
ambient medium density may be higher than previously thought.

The values of the parameters found in equation \ref{eq:results1}
imply that the comoving number density of the shocked plasma
(equation \ref{eq:n}) is 
\beq
\ba{lll}
n_p' & = & 5.0 \times 10^{14} \, (1+z)^{4/3} \,
d_{L,28.5}^{-2/3} \, \ee^{-7/3} \, \epl^{8/3} \, \eB^{-2/3} \nonumber
\\
& & \times
\theta_{p,0}^{-3} \, \Psi^{-7/3}(p) \, \alpha_1^{3/2} \,
\alpha_2^{-1/6} \, \alpha_3^{-2/3} \, \alpha_4^{-1/3} \, 
{\rm \,  cm^{-3}}.
\ea   
\label{eq:n2}
\eeq
For this value of the comoving number density, the plasma skin depth
is given by
\beq
\ba{lll}
\lambda \simeq { c \gamma_{\min}^{1/2} \over \omega_{pe}} & = & 0.2 \,
(1+z)^{-2/3} \, d_{L,28.5}^{1/3} \, \ee^{5/3} \, \epl^{-11/6} \,
\eB^{1/3} \nonumber \\
& & \times \theta_{p,0}^{2} \, \Psi^{5/3}(p) \, \alpha_1^{-3/4} \,
\alpha_2^{1/12} \, \alpha_3^{1/3} \, \alpha_4^{1/6} \, {\rm \, cm},  
\ea
\label{eq:skin_depth}
\eeq
where $\omega_{pe} = (4 \pi q^2 n_p' / m_e)^{1/2}$ is the plasma
frequency. 
This value of the skin depth implies that the magnetic field
decays on a characteristic length scale
\beq
\ba{lll}
{\Delta r'_B \over \lambda} & = & 2 \times 10^5 \, (1+z)^{-2/3} \,
d_{L,28.5}^{1/3} \, \ee^{-1/3} \, \epl^{1/6} \, \eB^{-2/3} \nonumber
\\
& & \times
\theta_{p,0}^{-1} \, \Psi^{-1/3}(p) \, \alpha_1^{-1/4} \,
\alpha_2^{-5/12} \, \alpha_3^{1/3} \, \alpha_4^{1/6}
\ea
\label{eq:dr_sd}
\eeq
skin depths. 
This decay length of the magnetic field is four orders of
magnitude shorter than the characteristic scale $\approx 10^9$ skin
depth assumed so far \citep{Piran05}. On the other hand, it is three  
orders of magnitude longer than the maximum length scale of magnetic
field generation that can be calculated using state of the art
numerical models \citep{Silva03, Fred04, Nishikawa05}.
The results obtained here are based on the interpretation of
GRB prompt emission spectra. They can therefore serve as a guideline
for the characteristic scale needed in future numerical models of
magnetic field generation.

The results presented in equations \ref{eq:results1}, \ref{eq:e_ssa2}
and \ref{eq:dr_sd} indicate that the value of $\epsilon_e$ should be
close to equipartition. The value of $\epsilon_B$ on the other hand,
is less constrained, and values as low as 1-2 orders of magnitude below
equipartition are consistent with the data (the stringent constraint
on the value of $\epsilon_B$ is obtained by the self absorption
energy, equation \ref{eq:e_ssa2}). 
The results presented in equation \ref{eq:results1} indicates that a low
value of $\epsilon_{pl}$ results in large emission radius and large
decay length of the magnetic field. Thus, low value of $\epsilon_{pl}$
implies that the model presented here can account for late time
flaring activities observed in many GRBs, that may originate from
shell collisions at large radii. A lower limit on the value of
$\epsilon_{pl}$ can be set by the requirement that the emission radius
is not larger than the transition radius to the self similar
expansion, $\sim 10^{16}$~cm which marks the beginning of the
afterglow emission phase. From this requirement, one obtains
$\epsilon_{pl} \gtrsim 10^{-2}$.

Generation of magnetic field and particle acceleration in shock waves
are most probably related issues \citep{Kazimura98, Silva03, Fred04,
HHFN04, Nishikawa05}. We therefore anticipate that the answers to the
theoretical questions raised by the model presented here, about the
requirement for high values of $\epsilon_e$ and $\epsilon_B$, the
uncertainty in the value of $\epsilon_{pl}$ and the characteristic
decay length of the magnetic field, are related to each other.

An underlying assumption in the calculations is that the values of
the free parameters are (approximately) constant inside the emitting
region. In reality, this of course may not be the case. We
introduced here a new length scale $\Delta r'_B$, characterizing
a length scale for the decay of $\epsilon_B$.  It can be argued that
within the context of this model the decay length of $\epsilon_e$ is
not shorter than $\Delta r'_B$. However, by defining $\Delta r'_B$ as
the shortest length within which both $\epsilon_e$ and $\epsilon_B$
maintain approximately constant values, the results presented here
hold.

The emission radius $r \approx 10^{12}$ cm found, implies that this
model can account for observed variability as short as $r/\Gamma^2 c
\approx 1$~ms. The observed GRB prompt emission spectra are usually
integrated over a much longer time scale, of few seconds. This can be
accounted for in our model, either by assuming low value of
$\epsilon_{pl}$, or by adopting the commonly used assumption
that the long duration emission is due to extended central engine
activity, which continuously produces new shock waves and refreshes
existing shock waves.

The results presented here are applicable to a large number of
astrophysical objects, in which magnetic field generation and particle
acceleration in shock waves are believed to play a major role.  Such
is the case for the study of afterglow emission from GRBs as well as
emission from supernovae remnants \citep[see, e.g., ][for the case of
SN1987A]{Chevalier92}. Additional astrophysical sources in which
strong shock waves and magnetic fields occur are active galactic
nuclei (AGNs) and jets in micro-quasars \citep{Fender03}. Current
observational status of these objects confines synchrotron emitting
regions only on a scale of $\sim 10^{13}$~cm \citep{DMR00}.  If the
length scale of the magnetic field inferred from observations in these
objects is found in the future to be similar to the value found here,
i.e., $\approx 10^5$ skin depths, this may serve as a strong hint
toward understanding magnetic field generation in shock waves.

\acknowledgements
AP wishes to thank Ralph Wijers, Peter M\'esz\'aros, Eli Waxman and
James Miller-Jones for useful discussions. This research was supported
by NWO grant 639.043.302, by the EU under RTN grant
HPRN-CT-2002-00294, and by NASA NNG05GB67G to BZ.

\appendix
\section{The dependence of the break energies on the power law index $p$ of
  the accelerated electrons}
\label{appendix}

\subsection{Electrons minimum Lorentz factor, $\gamma_{\min}$}
\label{psi_calc}

We assume that a fraction $\epsilon_{pl}$ of the electrons are
accelerated to a power law energy distribution $p$ above
$\gamma_{\min}$ and below $\gamma_{\max}$. While the initial value of
$\gamma_{\min}$ depends on the bulk Lorentz factor of the flow, a
quasi steady state of the electron distribution is formed, in which
the value of $\gamma_{\min}$ depends on the number and energy
densities of the acclerated particles. As this happens, the Lorentz
factor $\gamma_{\min}$ can be calculated given the number and energy
densities of the accelerated electrons. The electron energy
distribution is given by $dn/d\gamma = A \gamma^{-p}$, where $A$ is a
numerical constant. Integrating this function relates the values of
$\gamma_{\min}$ and $A$ to the number and energy densities of the
energetic electron component, $ \epsilon_{pl} n_{el} \simeq
\epsilon_{pl} n_p' = \int_{\gamma_{\min}}^{\gamma_{\max}} (dn/
d\gamma) d\gamma = A (1-p)^{-1} \left( \gM^{1-p} - \gm^{1-p} \right)$,
and
\beq
u_{el} \equiv \epsilon_e u' = m_e c^2 \int_{\gm}^{\gM} {dn \over
  d\gamma} \gamma d\gamma = m_e c^2 \times \left\{
\ba{ll}
A \log\left( {\gM \over \gm} \right)& (p=2), \nonumber \\
{A \over (2-p)} \left( \gM^{2-p} - \gm^{2-p} \right) & (p \neq 2).
\ea
\right.
\label{eq:A1}
\eeq
Dividing $u_{el}$ by $\epsilon_{pl} n_{el} m_e c^2$ eliminates $A$
from the equations,  
\beq
{u_{el} \over \epsilon_{pl} n_{el} m_e c^2}  =  \left\{
\ba{ll} 
\log\left( {\gM \over \gm} \right)
    \left( {1 \over \gm} - {1 \over \gM} \right)^{-1} & (p=2),
\nonumber \\ 
 \left( {1 - p \over 2-p}
\right) \left( {\gM^{2-p} - \gm^{2-p} \over \gM^{1-p} - \gm^{1-p} }
\right)   & (p \neq 2).
\ea
\right.
\eeq

We can now write the value of $\gamma_{\min}$ as $\gamma_{\min} =
[u_{el}/ (m_e c^2 \epsilon_{pl} n_{el})] (1 - \gamma_{\min}/\gamma_{\max})
\log(\gamma_{\max}/\gamma_{\min})^{-1} \times \Psi(p)$, where
$\Psi(p)$ is given by  
\beq
\Psi(p) =  \left\{
\ba{ll} 
1 & (p=2) \nonumber \\ 
\left( {p-2 \over p-1} \right) {\left[ \left( {\gm \over
        \gM}\right)^{p-1} -1 \right] \over \left[ \left( {\gm \over
        \gM}\right)^{p-2} -1 \right]} 
{\log\left( {\gM \over \gm} \right) \over \left[1-\left({\gm \over
    \gM}\right)\right]} 
& (p \neq 2). 
\ea
\right.
\label{eq:psi}
\eeq

\begin{figure}
\plotone{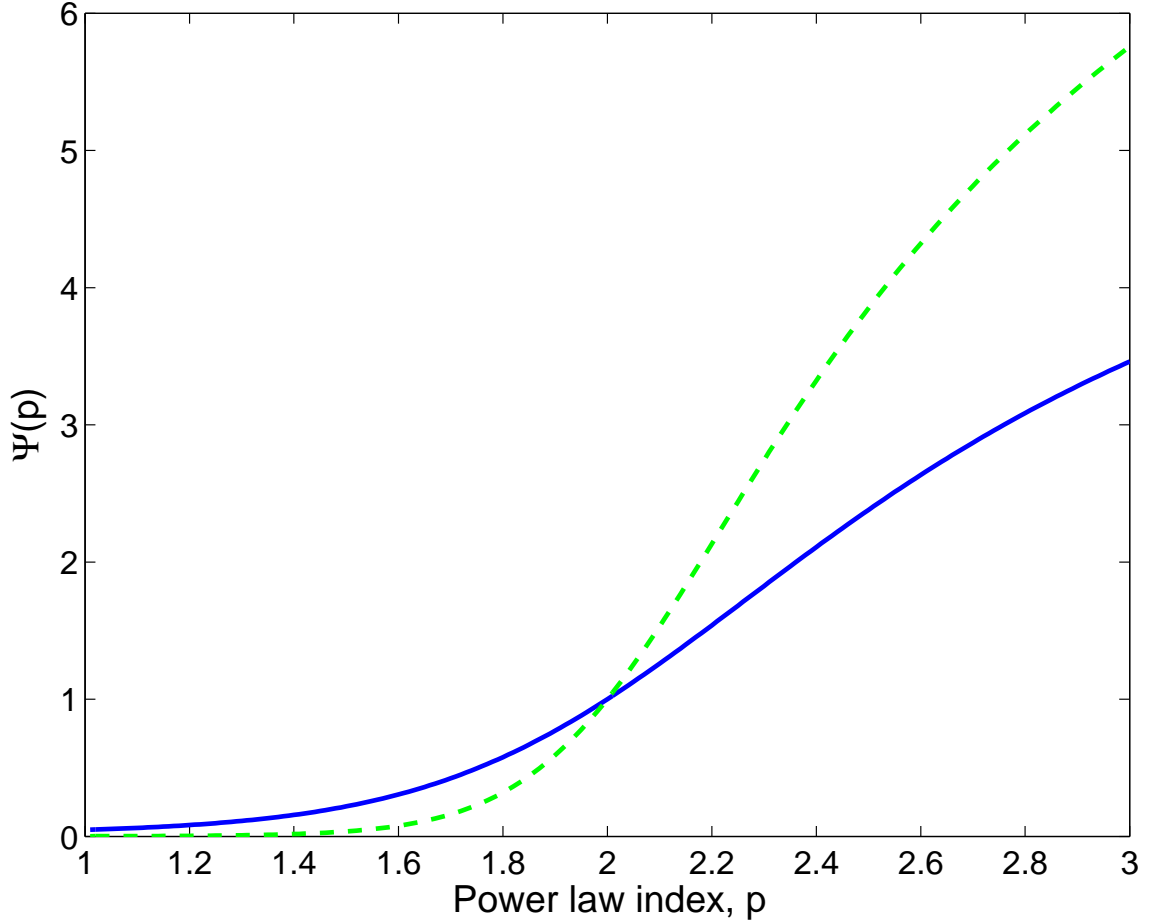}
\caption{Graph of the normalized function $\Psi(p)$ that determines
  the dependence of $\gamma_{\min}$ on the power law index $p$ of the
  accelerated electrons. Solid: $(\gamma_{\min}/\gamma_{\max}) =
  10^{-3}$, dash:  $(\gamma_{\min}/\gamma_{\max}) =
  10^{-5}$ (see equation \ref{eq:psi}). }
\label{fig:psi}
\end{figure}

The function $\Psi(p)$ is plotted in figure \ref{fig:psi} for two
representative values of ($\gamma_{\min}/\gamma_{\max}$)\footnote{The
value of $\gamma_{\max}$ can in principle be found from physical
constraints on the acceleration time. However, we find this way of
presentation much clearer.}.

\subsection{Self absorption energy, $\varepsilon_{ssa}^{ob.}$}
\label{chi_calc}

The synchrotron self absorption coefficient for a power law
distribution of electrons with power law index $p$ radiating in
magnetic field $B'$ is calculated using standard formula
\citep[e.g.,][]{Rybicki79},
\beq
\alpha_\nu = {\sqrt{27} q^4 \over 16 \pi^2 m_e^4 c^5} \left({3 q \over
  2 \pi m_e^3 c^5}\right)^{p/2-1} \left[{\Gamma\left({3p+2 \over 12}\right)
\Gamma\left({3p+22 \over 12}\right) \over \Gamma\left({2 \over
  3}\right)\Gamma\left({7 \over 3}\right)}\right]
\Gamma\left({2 \over3}\right)\Gamma\left({7 \over 3}\right) A
B'^{p/2+1} \nu^{-(p/2+2)},
\label{eq:alpha_nu}
\eeq
where the constant $A$ is calculated using equation \ref{eq:A1},
\beq
A = u_{el} \times \left\{
\ba{ll}
\log\left( {\gM \over \gm} \right)^{-1} & (p=2), \nonumber \\
{(p-2) \gm^{p-2} \over 1- \left( {\gm \over \gM} \right)^{p-2}} & (p \neq 2),
\ea
\right.
\eeq
which, upon insertion of $\gamma_{\min}$ can be written as 
\beq
A =  u_{el} \log\left( {\gM \over \gm} \right)^{-1}  \left({ u_{el}
  \over \epsilon_{pl} n_{el} m_e c^2}\right)^{p-2} \left[ 
  {1-(\gamma_{\min}/\gamma_{\max}) \over \log{\left({\gamma_{\max}
      \over \gamma_{\min}}\right)}}\right]^{p-2} \Psi^{p-2}(p) \xi(p),  
\label{eq:AA}
\eeq
where
\beq
\xi(p) = \left\{
\ba{ll} 
1 & (p=2),
\nonumber \\ 
 {(p-2) \log\left( {\gM \over \gm} \right)\over
1- \left({\gm \over \gM}\right)^{p-2}} & (p \neq 2).
\ea
\right.
\eeq

\begin{figure}
\plotone{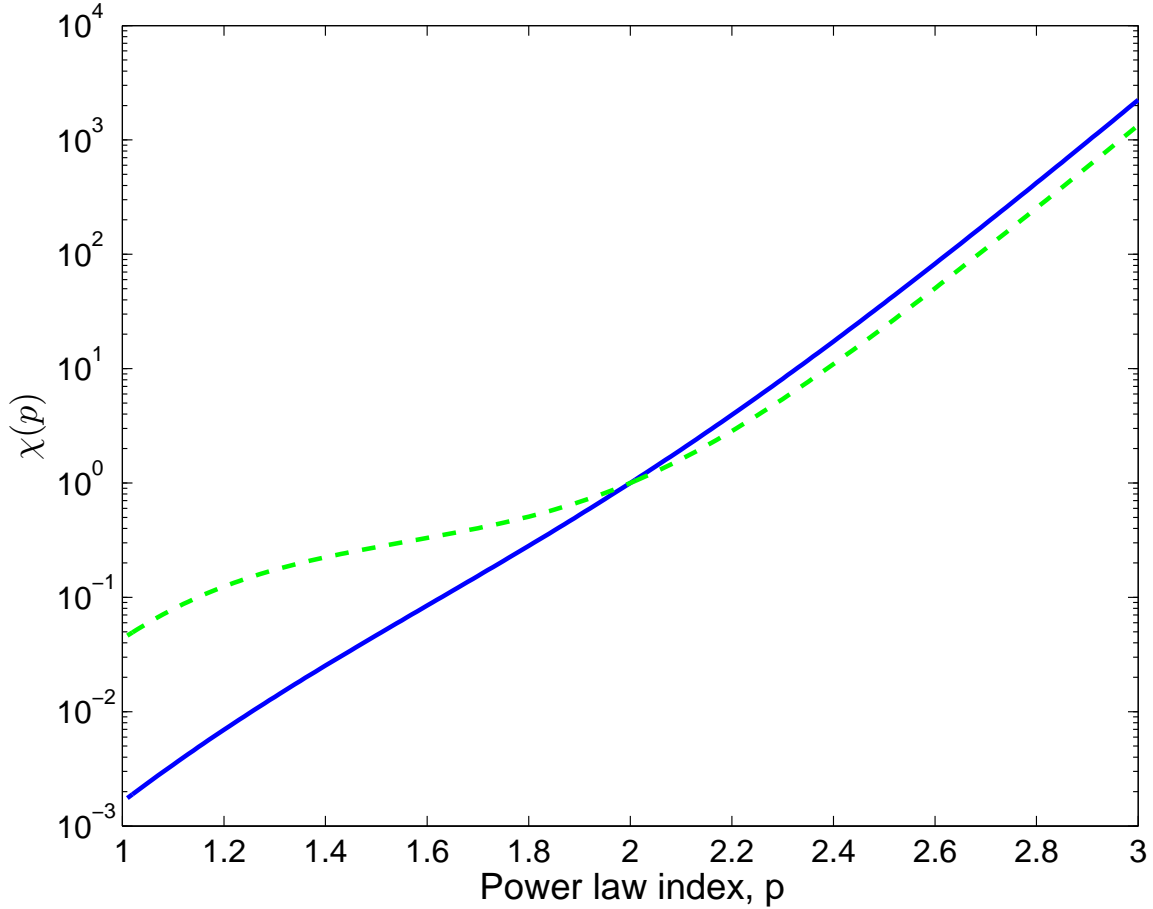}
\caption{Graph of the normalized function $\chi(p)$ that determines
  the dependence of $\epsilon_{ssa}^{ob.}$ on the power law index $p$ of the
  accelerated electrons. Solid: $(\gamma_{\min}/\gamma_{\max}) =
  10^{-3}$, dash:  $(\gamma_{\min}/\gamma_{\max}) =
  10^{-5}$ (see equation \ref{eq:e_ssa}). }
\label{fig:chi}
\end{figure}

Inserting the numerical values of the magnetic field and the peak
frequency $\nu_{peak} = \varepsilon_m^{ob.}/ \Gamma h$ (see equations
\ref{eq:B}, \ref{eq:em}) into the self absorption coefficient equation
\ref{eq:alpha_nu}, using the value of $A$ found in equation
\ref{eq:AA}, one obtains the synchrotron self absorption coefficient
at the peak frequency,
\beq
\ba{ll}
\alpha_{\nu_{peak}} = & 1.56 \times 10^{-11} \, \L^{1/2} \, \ri^{-1}
\Gi^{-1} \, \ee^{-5} \, \epl^{6} \, \eB^{-1/2} \, \theta_{p,0}^{-11/2} \,
\zeta_0^{1/2} \nonumber \\
& \times (3.0\times 10^{12})^{p/2-1} 
\left[{\Gamma\left({3p+2 \over 12}\right)
\Gamma\left({3p+22 \over 12}\right) \over \Gamma\left({2 \over
  3}\right)\Gamma\left({7 \over 3}\right)}\right] \xi(p) \Psi^{-6}(p)
\, {\rm cm^{-1}}.
\ea
\eeq
The self absorption optical depth $\tau_{\nu} = \Delta r'_B
\alpha_{\nu}$ is smaller than unity at $\nu = \nu_{peak}$. Since (by
demand) the electrons are in the slow cooling regime (i.e., $\gm \leq
\gamma_c$), the power radiated per unit energy below $\varepsilon_{m}
= \varepsilon_m^{ob.}/\Gamma$
is proportional to $(\varepsilon/\varepsilon_m)^{1/3}$, and the energy
below which the optical depth becomes greater than unity,
$\varepsilon_{ssa} = \varepsilon_m \tau_{\nu = \nu_{peak}}^{3/5}$, is 
\beq
\ba{ll}
\varepsilon_{ssa}^{ob.} = & {31 \over 1 + z} \, \L^{4/5} \, \ri^{-8/5}
\, \drb^{3/5} \, \Gi^{-3/5} \, \ee^{-1} \, \epl^{8/5} \, \eB^{1/5} \,
\theta_{p,0}^{-4/5} \, \zeta_0^{4/5} \nonumber \\
& \times (3.0\times 10^{12})^{(3/5)(p/2-1)} 
\left[{\Gamma\left({3p+2 \over 12}\right)
\Gamma\left({3p+22 \over 12}\right) \over \Gamma\left({2 \over
  3}\right)\Gamma\left({7 \over 3}\right)}\right]^{3/5} \xi^{3/5}(p)
\Psi^{-8/5}(p) \,  \eV
\ea
\label{eq:e_ssa3}
\eeq
(compare with equation \ref{eq:e_ssa}).
Therefore, the definition of the function $\chi(p)$ is 
\beq
\chi(p) \equiv (3.0\times 10^{12})^{(3/5)(p/2-1)} 
\left[{\Gamma\left({3p+2 \over 12}\right)
\Gamma\left({3p+22 \over 12}\right) \over \Gamma\left({2 \over
  3}\right)\Gamma\left({7 \over 3}\right)}\right]^{3/5} \xi^{3/5}(p)
\Psi^{-8/5}(p).
\label{eq:chi}
\eeq 
This function is plotted in figure \ref{fig:chi}.

\begin{figure}
\plotone{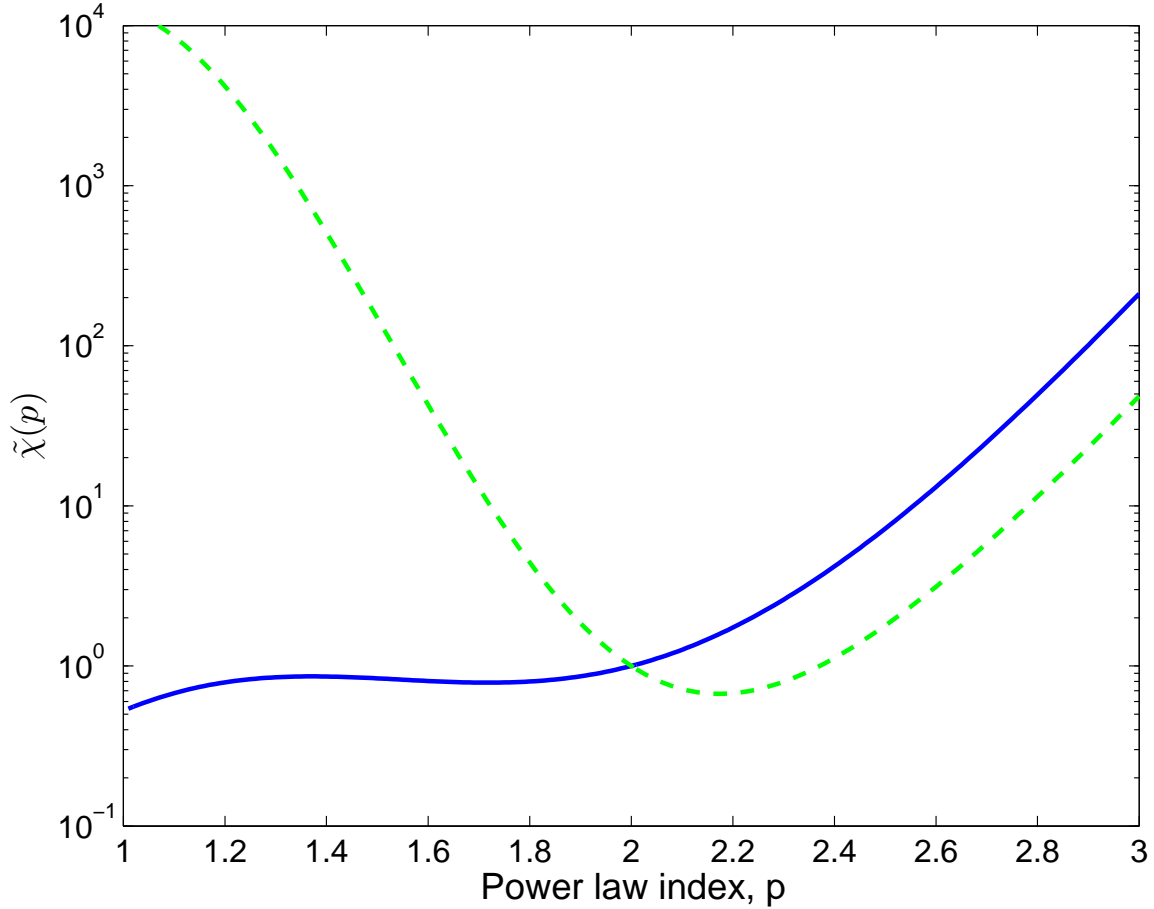}
\caption{Graph of the normalized function $\tilde{\chi}(p)$ that determines
  the dependence of $\varepsilon_{ssa}^{ob.}$ on the power law index
  $p$ of the accelerated electrons, for the parameters values derived
  from observations. Solid: $(\gamma_{\min}/\gamma_{\max}) =
  10^{-3}$, dash:  $(\gamma_{\min}/\gamma_{\max}) =
  10^{-5}$ (see equation \ref{eq:e_ssa2}). }
\label{fig:tild_chi}
\end{figure}

Inserting the parametric dependence on the value of $\Psi(p)$ of the
four parameters found in equation (\ref{eq:results1}) into equation
\ref{eq:e_ssa3}, leads to $\tilde{\chi}(p) = \chi(p) \times
\Psi^{-19/10}(p)$. Graph of this function appears in figure
\ref{fig:tild_chi}. Note that while $\chi(p)$ shows a very strong
dependence on the value of $p$, the function $\tilde{\chi}(p)$ varies
by a factor less than 4 in the range $2 \leq p \leq 2.4$.

\end{document}